# Otimizando A Alocação De Salas De Aula Com Foco Na Acessibilidade Para Pessoas Com Deficiência

**Resumo.** Este artigo aborda o desafio da alocação de salas de aula em instituições de ensino superior, enfatizando explicitamente a acessibilidade para Pessoas com Deficiência (PcDs). Empregando um estudo de caso do departamento de informática de uma universidade, o artigo propõe um modelo de otimização baseado em Programação Linear Inteira (PLI), sendo resolvido usando o *solver* Gurobi. O objetivo é minimizar o número de salas de aula usadas, priorizando a atribuição de alunos PcD às salas de aula do andar térreo para reduzir as barreiras de acessibilidade. O modelo é calibrado com um parâmetro de ponderação, α, que permite o equilíbrio entre a eficiência espacial e a promoção da acessibilidade. Os resultados experimentais indicam que o ajuste de α pode atingir um ponto de equilíbrio que melhora significativamente as práticas atuais de alocação manual, reduzindo o número de salas de aula necessárias e as penalidades de acessibilidade. As descobertas sugerem que os métodos de otimização podem melhorar a eficiência operacional em instituições acadêmicas, ao mesmo tempo, em que promovem um ambiente mais inclusivo para todos os alunos. Trabalhos futuros podem expandir a aplicação do modelo para outros departamentos e contextos e integrar critérios adicionais para desenvolver uma abordagem mais holística.

**Palavras-chave:** Alocação de salas; Acessibilidade; Otimização; Programação Linear Inteira; Inclusão.

# OPTIMIZING CLASSROOM ALLOCATION WITH A FOCUS ON ACCESSIBILITY FOR PEOPLE WITH DISABILITIES

**Abstract.** This paper addresses the challenge of classroom allocation in higher education institutions, with an explicit emphasis on accessibility for Persons with Disabilities (PwDs). Employing a case study of a university's computer science department, the paper proposes an Integer Linear Programming (ILP)-based optimization model, which is solved using the Gurobi solver. The objective is to minimize the number of classrooms used by prioritizing the assignment of PwD students to ground-floor classrooms to reduce accessibility barriers. The model is calibrated with a weighting parameter, α, that allows for a balance between spatial efficiency and promoting accessibility. Experimental results indicate that adjusting α can achieve a balance point that significantly improves current manual allocation practices, reducing the number of classrooms required and accessibility penalties. The findings suggest that optimization methods can improve operational efficiency in academic institutions while promoting a more inclusive environment for all students. Future work may expand the application of the model to other departments and contexts and integrate additional criteria to develop a more holistic approach.

**Keywords:** Room Allocation; Accessibility; Optimization; Integer Linear Programming; Inclusion.

## 1. INTRODUÇÃO

O Problema de Alocação de Salas (PAS) é um desafio comum em diversas áreas, especialmente educação, eventos e gestão de recursos. Otimizar esse processo requer a distribuição adequada de salas para diferentes atividades ou grupos, como alunos, palestrantes ou eventos, considerando critérios específicos, como capacidade da sala, proximidade entre as atividades e minimização de conflitos. Em instituições acadêmicas, a alocação eficiente de salas torna-se essencial para a organização e otimização de recursos para atender às necessidades de professores e alunos (CAVALCANTE et al., 2024). Considerar a acessibilidade para Pessoas com Deficiência (PcD) adiciona maior complexidade e relevância ao problema, ao implicar o comprometimento em criar um ambiente inclusivo que garanta o uso pleno das instalações para todos os membros da comunidade acadêmica. Esse contexto demanda incorporar critérios de acessibilidade física no processo de alocação de salas.

No contexto universitário, uma abordagem inclusiva é especialmente relevante, uma vez que a diversidade de alunos e suas necessidades específicas exigem um planejamento cuidadoso para garantir equidade e eficiência no uso dos espaços. Este cenário destaca a importância do desenvolvimento de modelos matemáticos e estratégias de otimização que integrem alocação eficiente de salas com requisitos fundamentais de acessibilidade para PcD . Este estudo explora a intersecção entre alocação de espaço e promoção da inclusão, propondo soluções inovadoras para um ambiente acadêmico mais acessível e justo.

Na prática universitária, a alocação de salas é uma atividade recorrente que ocorre a cada período acadêmico e envolve muitos alunos, incluindo alunos com deficiência. Além dos aspectos logísticos, as soluções propostas para alocação de salas precisam considerar a acessibilidade para permitir que todos os alunos participem totalmente das atividades acadêmicas. Em termos acadêmicos regulares, o processo de alocação de salas é particularmente exigente devido ao aumento do número de aulas e à complexidade do processo manual. Os regulamentos institucionais estabelecem a responsabilidade pela acessibilidade na alocação de salas para os gerentes acadêmicos. Dados esses desafios, a automação do processo de alocação surge como uma necessidade clara para garantir a integração efetiva das considerações de acessibilidade e o desenvolvimento de um ambiente acadêmico mais inclusivo.

Como solução para esses desafios, um sistema automatizado de alocação de salas oferece diversas vantagens em relação aos métodos manuais, como eficiência, precisão, redução de erros, flexibilidade, promoção da acessibilidade e geração de relatórios detalhados sobre as alocações realizadas. Esses benefícios contribuem para otimizar o processo de alocação, com redução de custos, melhoria da satisfação dos envolvidos e aumento da eficiência geral da instituição. Com base nessas considerações, este estudo aborda a modelagem e solução do problema de alocação de salas, enfatizando a distribuição de vagas para disciplinas de um curso específico em uma universidade. Na estrutura atual, a responsabilidade pela alocação de salas é dos gestores acadêmicos, e o processo, predominantemente manual, exige a consideração de diversos fatores, incluindo a identificação de alunos com deficiência em turmas específicas. O método manual tende a negligenciar a acessibilidade, o que resulta em retrabalho e desafios de gestão.

Para abordar esses desafios, este trabalho propõe um modelo matemático para otimização de processos, eliminando a necessidade de alocação manual e implementando uma abordagem automatizada para aumentar a eficiência e eficácia da tarefa. A alocação de salas universitárias com ênfase na acessibilidade para PcD representa uma área que não foi totalmente explorada em termos de automação, e este estudo visa preencher essa lacuna por meio de uma solução de otimização que priorize a acessibilidade.

Para atingir esses objetivos, a estrutura do trabalho é a seguinte: a Seção 2 inclui uma revisão da literatura relacionada ao problema abordado. A Seção 3 contém o estudo de caso usado para validar a modelagem proposta. A Seção 4 apresenta a modelagem do problema, seguida pela apresentação dos experimentos computacionais na Seção 5. Por fim, a Seção 6 apresenta as conclusões e sugestões para pesquisas futuras.

## 2. TRABALHOS RELACIONADOS

Vários pesquisadores têm proposto soluções para problemas de alocação em ambientes educacionais. Tomaz e Fonseca (2020) desenvolveram um modelo de correspondência de grafos e Programação Linear Inteira (PLI) para alocação de disciplinas, utilizando o solucionador Gurobi. O modelo é genérico e intuitivo, mas tem limitações na minimização de violações.

Couto et al. (2019) implementaram um algoritmo guloso e Simulated Annealing para maximizar a ocupação de salas no IFG - Campus Goiânia, mas enfrentaram crescimento exponencial do tempo de execução. Kripka et al. (2011) focaram na minimização de deslocamentos de alunos, utilizando Simulated Annealing, mas com restrições de tipos específicos de salas.

Silva et al. (2019) desenvolveram um modelo de PLI para alocação de salas na UFPB, incorporando preferências disciplinares, mas precisam melhorar a restrição de capacidade. De Freitas et al. (2019) aplicaram PLI e Gurobi para alocação de professores na Unicamp, com uma interface intuitiva, mas dependem de um parser e têm dificuldades na exportação de resultados.

Wodtke et al. (2022) propuseram um modelo de PLI para uma escola pública, reduzindo janelas de tempo, mas com limitações na exportação de resultados. Sales et al. (2015) desenvolveram um modelo de alocação de salas no Centro de Tecnologia da UFSM, focando em deslocamento e otimização do tempo de execução, mas com deficiências no tratamento da acessibilidade.

Prado e Souza (2014) utilizaram Simulated Annealing para alocação de salas, destacando-se pelo uso de um banco de dados e interface amigável, mas requer ajustes nas penalidades. Nunes et al. (2017) propuseram um modelo matemático para o PAS, reduzindo custos de deslocamento, mas necessitam de validação completa com dados reais. Jardim e Carvalho (2018) implementaram um software web para resolver o CAP, considerando todos os recursos das salas, mas necessitam de avaliação temporal.

2.1 Requisitos

A análise dos trabalhos relacionados destaca a necessidade de comparar requisitos obrigatórios (ROs) e não obrigatórios (NRs), além das variáveis fundamentais aplicadas na formatação dos modelos. ROs incluem restrições como número mínimo e máximo de dias entre reuniões, duração máxima de eventos, e alocação de disciplinas em salas apropriadas. NRs incluem recursos adicionais, proximidade ao bloco de aulas, e manutenção de professores na mesma sala.

2.2 Variáveis/Parâmetros

A revisão da literatura identificou dez variáveis principais recorrentes em estudos de otimização de alocação: eventos, professores, disciplinas, salas de aula, tempos, recursos, capacidades, ocupação, distância e acessibilidade. A análise quantitativa destaca a predominância das variáveis Assuntos (V3), Salas (V4) e Tempos (V5), seguidas da variável Capacidade (V7), sugerindo sua relevância fundamental para a formulação de soluções para problemas de alocação.

O modelo proposto nesta pesquisa estabelece alinhamento metodológico com estudos anteriores, utilizando parâmetros como V3, V4, V5, V7, V8 e V10, com foco prioritário na acessibilidade.

## 3. ESTUDO DE CASO

Esta pesquisa analisa o problema de alocação de salas em um curso de Ciência da Computação de uma universidade federal brasileira, visando melhorar o uso dos espaços e garantir acessibilidade para alunos com deficiência. O método adotado envolve modelagem matemática baseada em Programação Linear Inteira (ILP), que integra um sistema automatizado de alocação de salas.

A alocação de salas de aula em universidades representa um desafio complexo e recorrente, principalmente em contextos onde o número de alunos e cursos está em constante crescimento, assim como a diversidade de horários. Essa complexidade aumenta em cursos que utilizam uma infraestrutura mista composta por salas de aula convencionais e laboratórios. Para analisar esse cenário, este estudo se concentra nas disciplinas de um curso específico de Ciência da Computação de um departamento de ciência da computação, que possui uma variedade de espaços distribuídos em vários andares e blocos. A Tabela 1 apresenta uma amostra da capacidade e características das salas disponíveis e destaca o tipo de espaço e o andar de cada uma delas, fatores essenciais para uma alocação eficiente, principalmente quanto aos critérios de acessibilidade.

Tabela 1: Capacidade de Salas e Laboratórios Disponíveis para o Curso de Ciência da Computação

| Bloco | Sala | Tipo | Capacidade total | Andar |
|---|---|---|---|---|
| O | 101 | Sala Convencional | 40 | Térreo |
| B | 203 | Sala de estudo | 15 | 1º |
| B | 204 | Sala Convencional | 45 | 1º |
| C | 304 | Laboratório de Programação | 25 | 2º |
| E | 403 | Laboratório de Física | 20 | 3º |
| E | 504 | Sala Convencional | 40 | 4º |

Fonte: O próprio autor

Neste contexto, o Departamento de Ciência da Computação apoia vários cursos em ciências exatas e tecnologia e atende alunos de graduação e pós-graduação. Sua infraestrutura é composta por salas de aula e laboratórios de informática, compartilhados com outras unidades acadêmicas, especialmente em períodos de alta demanda. Este cenário apresenta desafios significativos para alocação eficiente de classes, particularmente aquelas que incluem alunos com necessidades específicas de acessibilidade.

### 3.1 PROBLEMA DE ALOCAÇÃO DE SALAS

Diante dessa realidade, o processo de alocação de salas de aula requer organização otimizada das turmas, visando garantir a acessibilidade e promover o uso racional dos recursos disponíveis. Para a execução desta pesquisa, o semestre letivo de 2023.1 foi selecionado como período de análise devido à sua relevância histórica e características singulares. Nesse período, as atividades presenciais retornaram após o fim da pandemia da COVID-19, o que possibilitou analisar as novas dinâmicas e os desafios institucionais na retomada do ensino presencial.

Em termos de funcionamento, as atividades acadêmicas da universidade ocorrem em três turnos distintos: manhã, tarde e noite, que compreendem 16 faixas horárias semanais. O calendário acadêmico estabelece os dias letivos como de segunda a sexta-feira, incluindo sábados em alguns cursos. O curso de Ciência da Computação, foco deste estudo, opera em turno integrado, com atividades nos períodos da tarde e noite, de segunda a sexta-feira.

### 3.2 MODELAGEM

O Problema de Alocação de Salas (PAS) apresenta variações com restrições e objetivos ambientais específicos. Este trabalho foi utilizado como referência para o período letivo de 2023.1, no qual o curso de Ciência da Computação ofereceu 59 aulas, o que resultou em 33 alocações diferentes. Essas alocações incluíram salas de aula, laboratórios, salas temporárias e aulas sem sala definida.

O processo de alocação de salas ocorre em três etapas principais: primeiro, o departamento responsável pré-oferece a turma e estabelece a relação Disciplina-Professor-Horário; segundo, o Centro faz a alocação com base nas informações da turma e no conjunto de salas priorizadas para cada curso; finalmente, o departamento recebe a alocação e disponibiliza as vagas para os alunos dos respectivos cursos.

O processo de alocação estabelece salas fixas para cada conjunto de classes (Disciplina-Horário-Professor). Este modelo é semelhante ao PAS tradicional, com a adição de critérios de acessibilidade. Com base no trabalho relacionado e nas características específicas do problema, os seguintes requisitos fundamentais foram estabelecidos para alocação de classes:

1. Cada disciplina deverá ser ministrada conforme a frequência prevista no horizonte de planejamento;
2. É proibida a atribuição simultânea de disciplinas diferentes na mesma sala;
3. É proibido atribuir um sujeito a várias salas ao mesmo tempo;
4. Disciplinas com necessidades laboratoriais específicas devem ocupar seus locais designados.

A modelagem matemática requer a definição de quatro elementos principais:

- Parâmetros: conjuntos de dados e informações básicas do modelo;
- Variáveis: elementos que representam escolhas possíveis;
- Restrições: limitações que condicionam as variáveis;
- Função objetivo: definição matemática dos objetivos do modelo.

### 3.2 1 PARÂMETROS

- $D$ : Conjunto de disciplinas; $T$ : Conjunto de horários; $S$ : Conjunto de salas;
- $PCD_d \in \{0, 1\}$: Indicador de presença de pessoa com deficiência na disciplina d;
- $Andar_s \in Z$: Número do andar onde a sala $s$ está localizada.

### 3.2.2 VARIÁVEIS DE DECISÃO

- $x_{dts} \in \{0, 1\}$: indicador de alocação, em que 1, a disciplina $d$ é alocada na sala $s$ no período $t$; e 0 caso contrário.

- $Y_s \in \{0, 1\}$: indicador de utilização do espaço, em que 1 indica que a sala $s$ está em uso; e 0 caso contrário.

### 3.2.3 RESTRIÇÕES

$$\sum_{t \in T} \sum_{s \in S} x_{dts} = 2 \qquad \forall\, d \in D \qquad (1)$$

Restrição (1): Garante que cada disciplina seja ensinada exatamente duas vezes por semana no horizonte de planejamento.

$$\sum_{d \in D} x_{dts} \leq 1 \qquad \forall\, t \in T,\, s \in S \qquad (2)$$

Restrição (2): Impede a alocação simultânea de vários sujeitos na mesma sala.

$$\sum_{s \in S} x_{dts} \leq 1 \qquad \forall\, d \in D,\, t \in T \qquad (3)$$

Restrição (3): Impede que um sujeito seja alocado em várias salas simultaneamente.

$$x_{dts} \leq y_s \qquad \forall\, d \in D,\, t \in T,\, s \in S \qquad (4)$$

Restrição (4): Vincula a utilização da sala à alocação de disciplinas, essencial para minimizar o número total de salas utilizadas.

### 3.2.4 FUNÇÃO OBJETIVO

$$Minimizar \quad (1 - \alpha) \sum_{s \in S} y_s + \alpha \times \sum_{d \in D} \sum_{t \in T} \sum_{s \in S} x_{dts} \times andar_s \times PCD_d \qquad (5)$$

A função objetivo (FO) deste modelo de otimização visa minimizar dois objetivos principais. O primeiro objetivo é reduzir o número de salas de aula utilizadas. O segundo objetivo envolve aplicar uma penalidade quando pessoas com deficiência (PcD) são alocadas em salas de aula que não estão localizadas no térreo (andar = 0). Esta penalidade aumenta proporcionalmente ao andar da sala de aula, sendo mais severa quanto mais alto for. Em outras palavras, a função busca alocar salas de aula com o menor número de turmas com PcDs em andares diferentes do térreo.

A função objetivo é complementada por um coeficiente de ponderação, representado por α, que permite flexibilidade na priorização de objetivos. O ajuste de α expressa a correlação: valores mais próximos de 0 enfatizam a minimização do número total de salas de aula, enquanto valores mais próximos de 1 priorizam alocar turmas com PCD em salas de aula do andar térreo.

# 4. EXPERIMENTOS COMPUTACIONAIS

Esta seção apresenta os resultados de experimentos computacionais que avaliam o desempenho do modelo de alocação de salas desenvolvido pelo Departamento de Ciência da Computação da Universidade em estudo. O principal objetivo do modelo é otimizar a alocação de salas, com prioridade para acessibilidade para Pessoas com Deficiência (PcD).

A implementação do modelo utilizou o Gurobi Solver e a linguagem de programação Python. Os testes, executados em um sistema com processador AMD Ryzen 5 5700X e 32 GB de RAM, comprovaram a eficiência do modelo, com tempos médios de execução de 0,03 segundos, apesar da complexidade das restrições e variáveis envolvidas.

## 4.1 CALIBRAÇÃO DO PARÂMETRO α

Por conta da natureza multi-objetivo do problema, realizaram-se experimentos de calibração para o parâmetro de ponderação α, cujos resultados constam na Tabela 2. O parâmetro $\alpha \in [0,1]$ estabelece a prioridade entre dois objetivos: a minimização do número de salas utilizadas (OBJ1) e a priorização da acessibilidade para Pessoas com Deficiência (PcD) (OBJ2). A coluna FO apresenta o valor da função objetivo resultante, a qual combina ambos os objetivos de modo ponderado. As colunas OBJ1 e OBJ2 contêm os valores específicos de cada objetivo, enquanto OBJ_W1 e OBJ_W2 representam as contribuições ponderadas na função objetivo total, após a aplicação do peso α.

Os resultados apresentados na Tabela 5 demonstram uma transição clara nas prioridades do modelo, pois α varia de 0 a 1. Com α = 0, o modelo foca exclusivamente em minimizar o número de cômodos (OBJ1 = 13), ignorando a penalidade associada a não alocação de PcD para andares inferiores (OBJ2 = 23). Para α = 1, o modelo prioriza a acessibilidade, aumentando substancialmente o número de cômodos (OBJ1 = 30) e minimizando a penalidade para PcD (OBJ2 = 8). Esse comportamento destaca a importância do equilíbrio, conforme observado em α = 0,5 quando ambos os objetivos são ponderados igualmente, resultando em um valor de função objetivo de 10,5, com 13 salas ocupadas e uma penalidade mínima de 8.

Além disso, a variação de α modifica o foco entre OBJ1 e OBJ2 e destaca a rigidez de certas restrições. Na situação em que α = 0, a solução minimiza o número de salas. Ainda assim, a penalidade de OBJ2 permanece alta, o que indica que o foco exclusivo em economia de espaço compromete significativamente a acessibilidade. Aumentar α causa uma rápida redução na penalidade de OBJ2 até α = 0,5; a partir desse ponto, os ganhos em acessibilidade diminuem enquanto o número de salas utilizadas aumenta rapidamente. Essa análise indica um ponto de equilíbrio natural em que qualquer aumento adicional na prioridade de acessibilidade produz custos significativamente maiores em termos de espaço utilizado.

Tabela 2: Resultados da calibração do parâmetro α.

| α | FO | OBJ1 | OBJ2 | OBJ_W1 | OBJ_W2 |
|---|---|---|---|---|---|
| 0 | 13 | 13 | 23 | 13 | 0 |
| 0.25 | 11.75 | 13 | 8 | 9.75 | 2 |
| 0.5 | 10.5 | 13 | 8 | 6.5 | 4 |
| 0.75 | 9.25 | 13 | 8 | 3.25 | 6 |
| 1 | 8 | 30 | 8 | 0 | 8 |

Em resumo, a análise dos resultados indica que as soluções para $0 < α < 1$ dominam as soluções em $α = 0$ e $α = 1$. Este resultado demonstra que, nas situações intermediárias, o modelo estabelece um compromisso prático e otimiza ambos os objetivos de forma equilibrada. A solução em $α = 0,5$ é particularmente relevante, pois apresenta um valor de função objetivo relativamente baixo (10,5) e equilibra o uso do espaço e a acessibilidade. A solução se destaca por minimizar os *trade-offs* entre os dois objetivos. Constitui uma solução Pareto-Ótima, na qual não é viável melhorar um dos objetivos sem comprometer o outro. Portanto, $α = 0,5$ foi estabelecido como o valor padrão para comparação com a solução de alocação adotada atualmente pelo departamento da universidade do estudo.

## 4.2 COMPARAÇÃO COM A SOLUÇÃO ATUAL

A análise comparativa entre o modelo proposto e a solução atual do Departamento de Ciência da Computação (ver Figura 6) revela aspectos significativos quanto à otimização da alocação de salas. Os resultados indicam que a alocação departamental utiliza 23 salas, representando 74% das 31 disponíveis. Para o cenário com $α = 0,5$, observa-se uma redução para 13 salas, resultando em um resultado equivalente à alocação com $α = 0$, cenário no qual o foco principal é a minimização. Esse resultado comprova a eficácia do coeficiente na redução do número de salas em uso. No caso de $α = 1$, quando não é considerada a minimização de salas, o modelo utiliza 30 salas, próximo ao total disponível.

Figura 6: Uso da sala.

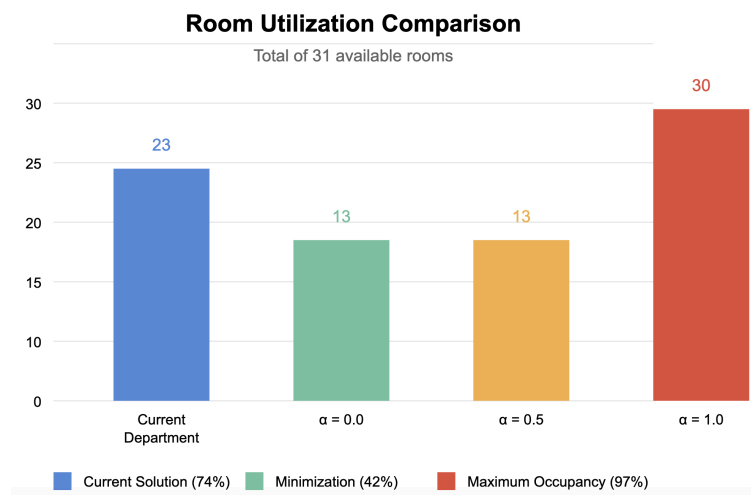

Fonte: O próprio autor

Além da análise quantitativa do uso da sala de aula, a avaliação de acessibilidade requer um estudo detalhado da distribuição das salas de aula em diferentes andares. Para essa análise, a Figura 7 apresenta um mapa de calor que ilustra a ocupação dos andares na alocação manual original e as soluções obtidas pelo modelo. Cada célula do mapa indica a proporção de sujeitos por andar, permitindo uma visualização clara das variações entre as diferentes abordagens implementadas.

Figura 7: Mapa de calor do piso.

| Floor | α = 0.0 | α = 0.25 | α = 0.50 | α = 0.75 | α = 1.0 |
|---|---|---|---|---|---|
| 4th | 80% | 70% | 60% | 45% | 30% |
| 3rd | 75% | 65% | 55% | 40% | 25% |
| 2nd | 45% | 50% | 60% | 70% | 75% |
| 1st | 30% | 45% | 65% | 80% | 85% |
| Ground | 20% | 40% | 70% | 85% | 95% |

**Occupancy Legend**
High (80-100%)   Medium (40-79%)   Low (0-39%)

Fonte: o próprio autor

Na Figura 7, o térreo (*ground*), essencial para promover a acessibilidade, apresenta baixa ocupação (20%) quando α=0,0, indicando que a prioridade inicial do modelo é minimizar o número de cômodos sem considerar a acessibilidade. À medida que α aumenta, a ocupação do térreo cresce consistentemente, chegando a 95% com α=1,0, quando a acessibilidade se torna o foco principal do modelo.

Nos andares superiores, o comportamento oposto é observado. O quarto andar, por exemplo, tem 80% de ocupação para α=0,0, mas essa taxa diminui gradativamente com o aumento de α, chegando a apenas 30% quando a acessibilidade é priorizada. Padrão semelhante ocorre em andares intermediários, como o terceiro andar, cuja ocupação diminui de 75% com α=0,0 para 25% com α=1,0.

Os andares intermediários (primeiro e segundo) apresentam flutuações mais equilibradas, destacando-se como uma transição entre o foco na minimização de cômodos e a priorização da acessibilidade. Para α=0,5, representando um equilíbrio entre os dois objetivos, o térreo atinge 70% de ocupação, enquanto o primeiro e o segundo andares se estabilizam em 65% e 60%, respectivamente. Este cenário indica um ponto de equilíbrio no qual a utilização dos cômodos é otimizada sem comprometer significativamente a acessibilidade.

Em resumo, os resultados mostram que o parâmetro α desempenha um papel crucial no ajuste das prioridades do modelo. Valores intermediários de α podem equilibrar a eficiência da ocupação da sala de aula e a inclusão de alunos com deficiência, enquanto valores extremos priorizam apenas um dos objetivos. Isso reforça a flexibilidade e a aplicabilidade do modelo em cenários com demandas específicas de inclusão e eficiência.

## 5. CONCLUSÕES

Este estudo apresentou uma solução para otimizar a alocação de salas de aula em uma universidade federal brasileira, priorizando alunos com deficiência (PcD) em um Departamento de Ciência da Computação. A metodologia incluiu coleta de dados reais, formulação matemática detalhada e implementação computacional com Programação Linear Inteira (PLI) e Python. Os resultados alcançados contribuíram para a pesquisa operacional e destacaram a importância de equilibrar múltiplos objetivos, como minimizar o uso de salas e maximizar a acessibilidade.O coeficiente de ponderação α foi essencial para estabelecer um equilíbrio efetivo entre esses objetivos. O modelo resolveu um problema prático, melhorou a distribuição de salas e ofereceu contribuições significativas para a gestão eficiente de recursos acadêmicos.

No entanto, o estudo apresentou limitações, como a restrição a um tipo específico de sala (laboratório de informática) e a segregação rigorosa de horários, necessitando de ajustes para maior flexibilidade. Em termos de acessibilidade, o modelo não incorporou critérios relacionados ao nível de acessibilidade dos edifícios ou à minimização do deslocamento dos alunos. Além disso, a alocação por turma e horário permitiu a distribuição de uma mesma turma em salas diferentes, podendo não atender às expectativas da gestão.

Para superar essas limitações, propõe-se a expansão do estudo para outros cursos e departamentos, a utilização de métodos heurísticos, o desenvolvimento de um sistema automatizado de extração de dados e a incorporação de métricas adicionais, como custos operacionais e sustentabilidade. Sugere-se também a criação de uma interface de usuário intuitiva para facilitar o acesso por administradores e gestores acadêmicos. Por fim, a implementação de heurísticas para casos grandes pode melhorar a eficiência e a escalabilidade do modelo, permitindo sua aplicação em contextos mais complexos e com maior número de variáveis (CLIMACO et al., 2022).

## 6. REFERÊNCIAS